\documentclass[useAMX,usenatbib,usegraphicx]{mn2e}

\usepackage{times}

\newcommand{\JHK}{{JHK_\mathrm{s}}}
\newcommand{\Ks}{{K_\mathrm{s}}}
\newcommand{\delmu}{{\Delta\mu_0}}

\title[P-L relations of type II Cepheids]
{Period-luminosity relations of type II Cepheids in the Magellanic Clouds}
\author[N. Matsunaga, M.W. Feast and I. Soszy\'{n}ski]
{Noriyuki Matsunaga$^{1}$\thanks{E-mail:matsunaga@ioa.s.u-tokyo.ac.jp},
Michael W. Feast$^{2,3}$ and Igor Soszy\'{n}ski$^{4}$\\
$^{1}$ Kiso Observatory, Institute of Astronomy, School of Science,
the University of Tokyo, 10762-30, Mitake, Kiso, Kiso, Nagano 397-0101, Japan\\
$^{2}$ South African Astronomical Observatory, PO Box 9, Observatory, 7935, South Africa\\
$^{3}$ Astrophysics, Cosmology and Gravitation Centre and 
Department of Astronomy, University of Cape Town, 
Rondebosch, 7701, South Africa\\
$^{4}$ Warsaw University Observatory, Al. Ujazdowskie 4, 00-478 Warszawa, Poland
}

\begin{document}

\date{Accepted 2010 November 29. Received 2010 November 24; in original form 2010 July 22}
\pagerange{\pageref{firstpage}--\pageref{lastpage}} \pubyear{2010}
\maketitle

\label{firstpage}

\begin{abstract}
Period-luminosity relations (PLRs) of type II Cepheids (T2Cs) in
the Small Magellanic Cloud are derived
based on OGLE-III, IRSF/SIRIUS and other data, and these are
compared with results
for the Large Magellanic Cloud and Galactic globular clusters.
Evidence is found for a change of the PLR slopes from system to system.
Treating the longer period T2Cs (W~Vir stars)
separately gives an SMC-LMC modulus difference of $0.39\pm 0.05$~mag without
any metallicity corrections being applied. This agrees well with
the difference in moduli based on different distance indicators,
in particular the PLRs of classical Cepheids.
The shorter period T2Cs (BL~Her stars) give a smaller
SMC-LMC difference suggesting that their absolute magnitudes
might be affected either by metallicity or 
by age effects. It is shown that the frequency distribution
of T2C periods also changes from system to system.
\end{abstract}


\section{Introduction}
\label{sec:Intro}

Cepheids are pulsating stars with periods between one day
and about one hundred days. Although there was originally some confusion,
see the review by \citet{Fernie-1969}, it is now believed that Cepheids
can be grouped into two distinct classes: classical Cepheids and type II
Cepheids (T2Cs, hereafter). The former are intermediate mass stars
(4--10~$M_\odot$), while
the latters are lower-mass stars ($\sim 1~M_\odot$)
belonging to disc and halo populations 
(\citealt{Wallerstein-2002}; \citealt{Sandage-2006}).

T2Cs are conventionally divided into
three period groups, BL~Her stars (henceforth BL) at short periods,
W~Vir stars (WV) at
intermediate periods and RV~Tau stars (RV) at the longest periods.
According to Gingold~(\citeyear{Gingold-1976}, \citeyear{Gingold-1985}),
the BLs are stars evolving from the
horizontal branch to the Asymptotic Giant Branch (AGB); the WVs are stars
that along the AGB cross the Cepheid instability strip due to excursions
towards higher effective temperatures; the RVs are stars that after
their AGB phase are moving towards the white dwarf cooling sequence.
However, some later evolutionary tracks (e.g.~\citealt{Pietrinferni-2006})
do not show the excursions,
and the precise evolution state
of T2Cs remains unclear.
We follow the division by \citet{Soszynski-2008b}
and adopt thresholds of 4~d and 20~d to divide the three groups.
The RV stars tend to show alternating
deep and shallow minima (and this is often taken as a defining characteristic),
but the single period of the RVs will be used in this paper.  
In addition to these three groups, \citet{Soszynski-2008b} 
established a new group of T2Cs, peculiar W~Vir (pW) stars.
They have distinctive light curves and tend to be  brighter than normal
T2Cs of the same period.
It is suggested that many, perhaps all, of them are binaries, 
but their nature remains uncertain.
The bright Galactic T2C $\kappa$~Pav appears to be a pW star
(Matsunaga, Feast \& Menzies, 2009, hereafter M09).

\citeauthor{Matsunaga-2006}~(\citeyear{Matsunaga-2006}, hereafter M06)
discovered a tight PLR of T2Cs in globular clusters
based on infrared photometry for 46 T2Cs. 
\citet{Feast-2008} used pulsation parallaxes of nearby T2Cs
to calibrate this cluster PLR and to discuss the distances
of the Large Magellanic Cloud (LMC) and the Galactic Centre.
The Optical Gravitational Lensing Experiment (OGLE-III)
have significantly increased samples of T2Cs in the Magellanic Clouds.
\citet{Soszynski-2008b} found 197 T2Cs in the LMC.
M09 investigated their near-infrared (near-IR) nature and confirmed that
BL and WV stars follow a tight PLR
like that of T2Cs in globular clusters, whereas 
RV and pW stars show a large scatter and are systematically brighter than
the PLR for the BL and WVs.
The calibration by \citet{Feast-2008} then leads to a distance modulus
of the LMC of $18.46\pm 0.05$~mag.
These investigations have established the T2Cs
as a promising distance indicator.

The aim of the present paper is to compare 
the T2C PLRs in the Small Magellanic Cloud (SMC), LMC and globular clusters.
This is important for a further study of these relations and
their possible dependence on metallicity and other factors.
The OGLE-III survey discovered 43 T2Cs in the SMC
(\citealt{Soszynski-2010b}, S10 hereafter).
In the following, we first collect near-IR $JHK$ magnitudes
of SMC T2Cs
(Section~\ref{sec:Photometry}) and investigate their PLRs
(Section~\ref{sec:PLR}).
The reddening-free PLR in $VI$ is also discussed.
In parallel with our work,
\citeauthor{Ciechanowska-2010}~(\citeyear{Ciechanowska-2010}, C10)
have obtained independent $J$ and $K$ for a subset of the SMC T2Cs
and their data is combined with ours in some of the discussions. 
We find evidence that the BL and WV stars need to be discussed independently.
In Section~\ref{sec:delmu} the apparent differences in distance moduli of
the LMC and SMC are derived for these stars and compared with the differences
obtained from other objects.
The data used for classical Cepheids are discussed in an Appendix.
In addition, the colours and period distributions
of T2Cs in various systems are compared and contrasted
in Sections~\ref{sec:evolution}.
Section~\ref{sec:Conclusion} summarizes the present work.

\section{Infrared Photometry}
\label{sec:Photometry}

S10 catalogued 43~T2Cs in the SMC:
17~BLs, 10~WVs, 7~pWs and 9~RVs.
We searched for near-IR counterparts of these stars in the catalogue
of \citet{Kato-2007}.
This point-source catalogue is based
on simultaneous images in $\JHK$ obtained with the 1.4-m
Infrared Survey Facility (IRSF) located at South African Astronomical
Observatory (SAAO), Sutherland, South Africa.
The catalogue covers
$\rm 40~deg^{2}$ of the LMC, $\rm 11~deg^{2}$ of the SMC and
$\rm 4~deg^{2}$ of the Magellanic Bridge.
The catalogue extends to fainter magnitudes and has higher resolution than
the point-source catalogue of the Two-Micron All-Sky Survey
(2MASS; \citealt{Skrutskie-2006}) in the regions of the Magellanic Clouds.
The 10~$\sigma$ limiting magnitudes of the survey in the SMC are 18.9, 17.9 and
16.9~mag for $J$, $H$ and $\Ks$.
These limits are fainter than those for the LMC since the crowding 
effect is less \citep{Kato-2007}.

Matches between the OGLE-III and IRSF catalogues were found
for 39 S10 T2C sources
with a tolerance of 0.5$^{\prime\prime}$ (Table~\ref{tab:catalogue}).
The differences in coordinates are small between the catalogues
with standard deviations of less than 0.1$^{\prime\prime}$
in both RA and Dec.
Four OGLE-III sources (IDs in S10: \#5, \#37, \#42 and \#43) 
are located outside the IRSF survey field.
In the case of two matched objects, \#6 and \#22,
$\Ks$-band magnitudes are missing because
of the faintness of these stars, 
both of which are BL stars with relatively
short periods ($P<1.5$~d).
Although about half
of the BL sample have $\Ks$ fainter than the 10~$\sigma$ limiting 
magnitude, their $J$ and $H$ magnitudes are brighter than the limits.

\begin{table*}
\begin{minipage}{0.95\hsize}
\caption{
The catalogue of OGLE-III T2Cs in the SMC
with IRSF counterparts.  Modified Julian Dates (MJD),
pulsation phase of the observations,
$\JHK$ magnitudes and their errors are listed for each IRSF measurement
as well as the OGLE-IDs, types and periods.
Shifts for the phase corrections ($\delta_{\phi}$)
obtained from the $I$-band light curves are also listed.
Two sources, \#28 and \#35, are listed twice because they are located
in overlapping fields of the IRSF survey.
\label{tab:catalogue}}
\begin{center}
\begin{tabular}{ccccccccccccr}
\hline
 OGLE-ID & Type & $\log P$ & \multicolumn{9}{c}{IRSF counterpart} & \multicolumn{1}{c}{$\delta _{\phi}$} \\
 & & & IRSF-Field & MJD(obs) & Phase & $J$ & $E_J$ & $H$ & $E_H$ & $\Ks$ & $E_\Ks$ & \\
\hline
1   & pWVir & 1.07441 & SMC0031-7350F & 52847.100 & 0.944 & 14.32 & 0.01 & 14.01 & 0.01 & 14.00 & 0.01 & $ 0.076$ \\
2   & BLHer & 0.13741 & SMC0037-7250C & 52518.027 & 0.115 & 17.84 & 0.04 & 17.56 & 0.08 & 17.65 & 0.14 & $ 0.145$ \\
3   & WVir  & 0.63947 & SMC0036-7350G & 52849.096 & 0.677 & 17.27 & 0.03 & 16.95 & 0.04 & 17.06 & 0.14 & $-0.160$ \\
4   & WVir  & 0.81514 & SMC0037-7310A & 52517.858 & 0.911 & 16.16 & 0.04 & 15.80 & 0.04 & 15.69 & 0.05 & $ 0.025$ \\
6   & BLHer & 0.09188 & SMC0041-7330A & 52494.999 & 0.263 & 18.06 & 0.05 & 17.93 & 0.11 &    -- &   -- & $ 0.094$ \\
7   & RVTau & 1.49081 & SMC0041-7310D & 52496.056 & 0.372 & 13.10 & 0.01 & 12.74 & 0.01 & 12.62 & 0.02 & $-0.090$ \\
8   & BLHer & 0.17312 & SMC0046-7330I & 53413.759 & 0.552 & 17.60 & 0.05 & 17.03 & 0.06 & 17.05 & 0.24 & $-0.086$ \\
9   & BLHer & 0.47291 & SMC0046-7250C & 52457.193 & 0.867 & 17.17 & 0.03 & 16.81 & 0.03 & 16.73 & 0.07 & $ 0.118$ \\
10  & pWVir & 1.24256 & SMC0046-7330E & 53301.789 & 0.454 & 13.83 & 0.02 & 13.52 & 0.02 & 13.45 & 0.02 & $-0.025$ \\
11  & pWVir & 0.99675 & SMC0046-7330E & 53301.789 & 0.451 & 14.34 & 0.01 & 14.03 & 0.01 & 13.96 & 0.02 & $-0.052$ \\
12  & RVTau & 1.46566 & SMC0046-7310B & 53397.808 & 0.477 & 14.70 & 0.01 & 14.28 & 0.01 & 14.26 & 0.02 & $-0.044$ \\
13  & WVir  & 1.14019 & SMC0046-7250B & 52457.177 & 0.216 & 15.73 & 0.01 & 15.29 & 0.02 & 15.18 & 0.02 & $ 0.004$ \\
14  & WVir  & 1.14234 & SMC0050-7330I & 52517.837 & 0.622 & 15.98 & 0.02 & 15.56 & 0.02 & 15.50 & 0.03 & $-0.107$ \\
15  & BLHer & 0.40986 & SMC0050-7310E & 52517.914 & 0.223 & 16.59 & 0.02 & 16.41 & 0.03 & 16.40 & 0.06 & $ 0.037$ \\
16  & BLHer & 0.32494 & SMC0050-7250H & 52518.878 & 0.222 & 17.51 & 0.03 & 17.16 & 0.06 & 17.06 & 0.12 & $ 0.084$ \\
17  & BLHer & 0.11371 & SMC0051-7130B & 52534.061 & 0.483 & 17.60 & 0.04 & 17.39 & 0.08 & 17.04 & 0.14 & $-0.067$ \\
18  & RVTau & 1.59681 & SMC0050-7350G & 52894.012 & 0.409 & 14.36 & 0.02 & 13.75 & 0.01 & 13.12 & 0.01 & $-0.301$ \\
19  & RVTau & 1.61185 & SMC0055-7330C & 52868.173 & 0.867 & 13.93 & 0.02 & 13.71 & 0.01 & 13.62 & 0.02 & $ 0.043$ \\
20  & RVTau & 1.70435 & SMC0055-7230C & 52518.940 & 0.088 & 14.11 & 0.02 & 13.85 & 0.02 & 13.74 & 0.02 & $ 0.226$ \\
21  & BLHer & 0.36420 & SMC0055-7250F & 52842.164 & 0.890 & 17.77 & 0.04 & 17.43 & 0.10 & 17.24 & 0.14 & $ 0.029$ \\
22  & BLHer & 0.16747 & SMC0055-7350E & 52950.854 & 0.627 & 18.64 & 0.10 & 18.19 & 0.20 &    -- &   -- & $-0.222$ \\
23  & pWVir & 1.24737 & SMC0055-7310E & 52837.125 & 0.356 & 14.68 & 0.01 & 14.38 & 0.01 & 14.23 & 0.02 & $-0.034$ \\
24  & RVTau & 1.64307 & SMC0055-7330H & 52835.157 & 0.504 & 14.11 & 0.01 & 13.74 & 0.01 & 13.66 & 0.02 & $-0.065$ \\
25  & pWVir & 1.15140 & SMC0055-7310D & 52837.106 & 0.342 & 15.28 & 0.02 & 14.94 & 0.01 & 14.82 & 0.02 & $-0.050$ \\
26  & BLHer & 0.23168 & SMC0059-7210F & 53300.954 & 0.988 & 17.38 & 0.03 & 17.19 & 0.06 & 16.93 & 0.14 & $ 0.055$ \\
27  & BLHer & 0.18801 & SMC0100-7330F & 52844.081 & 0.752 & 18.11 & 0.04 & 17.81 & 0.08 & 17.62 & 0.15 & $-0.067$ \\
28  & pWVir & 1.18368 & SMC0055-7330D & 52835.090 & 0.595 & 14.58 & 0.01 & 14.19 & 0.01 & 14.08 & 0.02 & $-0.102$ \\
28  & pWVir & 1.18368 & SMC0100-7330F & 52844.081 & 0.184 & 14.32 & 0.01 & 13.94 & 0.01 & 13.81 & 0.01 & $ 0.034$ \\
29  & RVTau & 1.52733 & SMC0059-7210C & 52530.910 & 0.330 & 13.07 & 0.01 & 12.64 & 0.01 & 12.54 & 0.01 & $-0.084$ \\
30  & BLHer & 0.53006 & SMC0100-7310I & 52847.141 & 0.310 & 16.43 & 0.01 & 16.12 & 0.03 & 16.02 & 0.05 & $-0.000$ \\
31  & WVir  & 0.89737 & SMC0059-7130H & 52530.855 & 0.241 & 16.36 & 0.02 & 15.96 & 0.03 & 15.89 & 0.05 & $ 0.003$ \\
32  & WVir  & 1.15372 & SMC0100-7330B & 52844.177 & 0.161 & 15.50 & 0.02 & 15.21 & 0.01 & 15.07 & 0.02 & $ 0.093$ \\
33  & BLHer & 0.27362 & SMC0059-7230E & 53300.908 & 0.738 & 17.29 & 0.02 & 16.84 & 0.04 & 16.78 & 0.11 & $-0.232$ \\
34  & WVir  & 1.30364 & SMC0059-7210B & 52530.895 & 0.047 & 14.84 & 0.02 & 14.49 & 0.01 & 14.38 & 0.02 & $ 0.302$ \\
35  & WVir  & 1.23506 & SMC0100-7350D & 53355.802 & 0.706 & 15.46 & 0.02 & 15.19 & 0.02 & 15.16 & 0.04 & $-0.069$ \\
35  & WVir  & 1.23506 & SMC0100-7350E & 52956.766 & 0.481 & 15.86 & 0.02 & 15.49 & 0.02 & 15.44 & 0.05 & $-0.548$ \\
36  & BLHer & 0.03809 & SMC0104-7310F & 52869.164 & 0.853 & 17.52 & 0.04 & 17.43 & 0.09 & 17.23 & 0.13 & $ 0.017$ \\
38  & pWVir & 0.64778 & SMC0104-7330B & 52849.160 & 0.297 & 15.82 & 0.02 & 15.48 & 0.02 & 15.37 & 0.03 & $ 0.020$ \\
39  & BLHer & 0.27590 & SMC0109-7310F & 52956.779 & 0.398 & 17.31 & 0.03 & 16.98 & 0.07 & 17.15 & 0.24 & $-0.037$ \\
40  & WVir  & 1.20712 & SMC0108-7150D & 53315.777 & 0.985 & 15.04 & 0.01 & 14.71 & 0.01 & 14.60 & 0.04 & $ 0.367$ \\
41  & RVTau & 1.46417 & SMC0113-7310B & 52960.863 & 0.944 & 14.78 & 0.01 & 14.49 & 0.01 & 14.39 & 0.02 & $ 0.060$ \\
\hline
\end{tabular}
\end{center}
\end{minipage}
\end{table*}

\begin{figure}
\begin{minipage}{80mm}
\begin{center}
\includegraphics[clip,width=0.98\hsize]{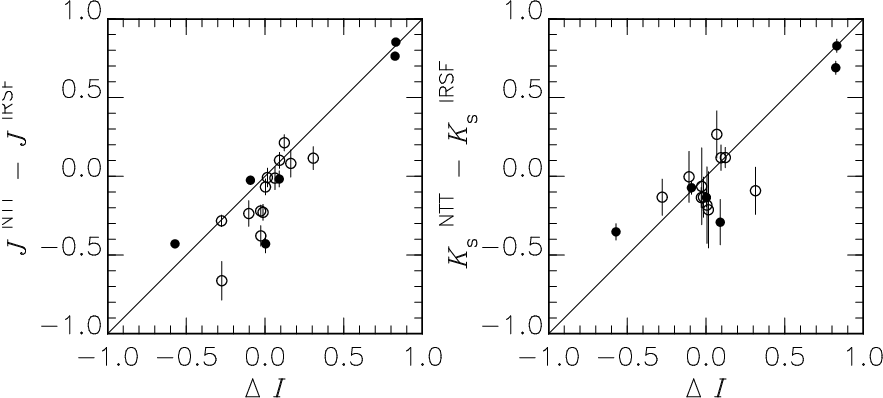}
\end{center}
\caption{
Differences between the IRSF photometry and the NTT photometry
(C10) are plotted against
the predicted differences from the $I$-band light curves.
BL stars are indicated by open circles and WV stars 
by filled circles.
\label{fig:comp_IRSF_Cie10}}
\end{minipage}
\end{figure}

We have compared our photometry with that obtained by
C10, who used the SOFI infrared camera attached to
the ESO New Technology Telescope (NTT) to obtain photometry in $J$ and $K$
for 19~OGLE-III T2Cs in the SMC (14~BLs and 5~WVs).
C10 observed three of them twice and the others once.
Excluding three objects not included in the IRSF survey,
there are 19~measurements for the T2Cs common with ours.
Their magnitudes are listed by C10 in the 2MASS system.
They were transformed into the IRSF/SIRIUS system \citep{Kato-2007}
before compared with our photometry.
Additionally, the OGLE-III (S10) periods and epochs of maximum $I$ were used
to calculate the phases for both the NTT observations and the IRSF ones
and to estimate the difference $\Delta I$ between the two phases.
Fig.~\ref{fig:comp_IRSF_Cie10} plots the difference between
their photometry and ours against the variation predicted
from the $I$-band light curves.
The plots show a reasonable correlation which indicates that
the NTT photometry is in agreement with our IRSF photometry.
It also supports the assumption that the $I$-band light curves reasonably
predict the near-infrared variations as suggested in M09,
so that we adopt the phase correction used by M09 to estimate mean
magnitudes\footnote{Phase-corrected data are used throughout this paper although
C10 found that this did not reduce the
scatter for their sample of stars. None
of the conclusions of the paper are significantly affected if uncorrected
data are used.}.

\section{Period-luminosity Relations of type II Cepheids}
\label{sec:PLR}

\subsection{PLRs of the SMC T2Cs}
\label{sec:SMC_T2C_PLR}

\begin{figure}
\begin{minipage}{85mm}
\begin{center}
\includegraphics[clip,width=0.85\hsize]{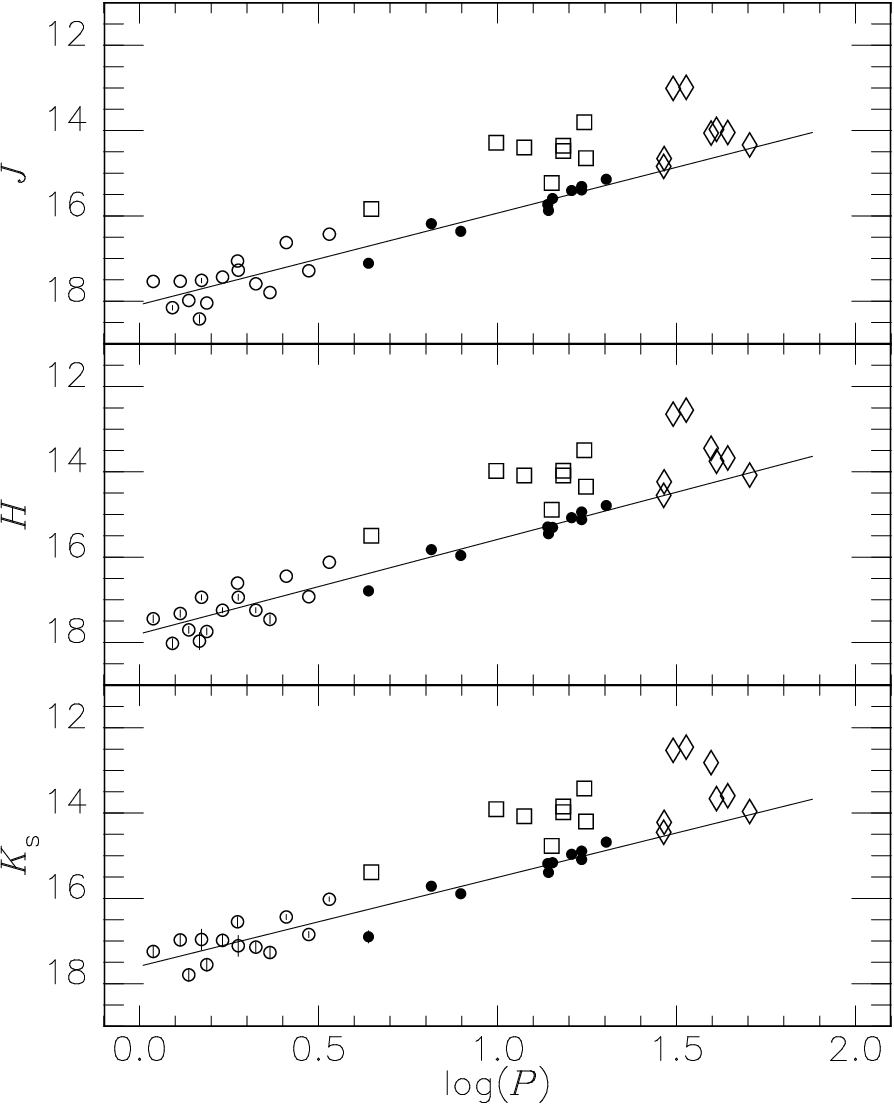}
\end{center}
\caption{Period-magnitude relations of T2Cs in the SMC.
The phase-corrected $\JHK$ magnitudes are plotted against periods.
Open circles indicates BL stars and filled circles WV stars.
The lines show the PLRs for the combined sample of BLs and WVs.
RV and pW stars are indicated by diamond symbols and squares.
Error bars are drawn if the photometric errors are larger than 0.05~mag.
\label{fig:PLR}}
\end{minipage}
\end{figure}

Fig.~\ref{fig:PLR} shows the phase-corrected $\JHK$ data plotted
against $\log P$ for the SMC T2Cs based on our data alone.
The BL and WV stars (the filled circles)
show well defined PLRs in all the $\JHK$ bands.
In contrast, pWs (squares) and RVs (open circles) show a large scatter
and are brighter than the PLRs of the BL and WV stars.
These objects are discussed in Section~\ref{sec:evolution}.

PLRs were calculated for BLs and WVs alone and
for the combined sample of BL and WV stars. 
We also derived reddening-free PLRs for the S10 $VI$ 
data using the indices, $W_i (VI) = I-R^{VI}_i (V-I)$,
where $R^{VI}_1=1.55$ and $R^{VI}_2=1.45$ corresponding to two reddening laws
as in M09. Table~\ref{tab:PLRs_T2C} also gives similar 
relations for the T2Cs in the LMC and those in globular clusters,
using the data in M09 and M06.
There is not sufficient data to derive PLRs in $VI$ for the globular cluster
objects.

\begin{table*}
\begin{minipage}{0.95\hsize}
\caption{
Period-luminosity relations of type II Cepheids.
The slope, zero-point near the median
period, residual scatter $\sigma$
and the sample size $N$ are listed for each group.
The relations in $\JHK$ bands are derived using only the IRSF data.
Those indicated as $J^\dagger$ and $\Ks^\dagger$ are
obtained from the combined IRSF and NTT (C10) datasets.
The reddening-free indices $W_{i}(VI)$ are defined in the text.
For the LMC/SMC $\JHK$ observations, phase corrections have been applied.
Mean magnitudes were obtained from repeated photometry for
globular clusters objects.
The PLRs for globular clusters are from
\citet{Matsunaga-2006} and those for the LMC from \citet{Matsunaga-2009}.
\label{tab:PLRs_T2C}}
\begin{center}
\begin{tabular}{ccccccc}
\hline
Band & Group   & Type & Slope & Zero (@~$\log P_0$) & $\sigma$ & $N$ \\
\hline
$J$        &SMC&BL   &$-2.545\pm 0.764$&$17.393\pm 0.112$ (@~0.3)&0.41& 15\\ 
$J$        &SMC&WV   &$-2.667\pm 0.246$&$15.483\pm 0.059$ (@~1.2)&0.16& 10\\ 
$J$        &SMC&BL+WV&$-2.147\pm 0.154$&$15.506\pm 0.116$ (@~1.2)&0.34& 25\\ 
$H$        &SMC&BL   &$-2.765\pm 0.731$&$17.080\pm 0.108$ (@~0.3)&0.40& 15\\ 
$H$        &SMC&WV   &$-2.655\pm 0.236$&$15.129\pm 0.057$ (@~1.2)&0.15& 10\\ 
$H$        &SMC&BL+WV&$-2.214\pm 0.148$&$15.141\pm 0.112$ (@~1.2)&0.32& 25\\ 
$\Ks$      &SMC&BL   &$-2.096\pm 0.732$&$16.933\pm 0.104$ (@~0.3)&0.37& 13\\ 
$\Ks$      &SMC&WV   &$-2.844\pm 0.305$&$15.038\pm 0.073$ (@~1.2)&0.20& 10\\ 
$\Ks$      &SMC&BL+WV&$-2.082\pm 0.151$&$15.091\pm 0.109$ (@~1.2)&0.32& 23\\ 
$J^\dagger$ &SMC&BL   &$-2.690\pm 0.488$&$17.325\pm 0.069$ (@~0.3)&0.36& 31\\
$J^\dagger$ &SMC&   WV&$-2.460\pm 0.277$&$15.486\pm 0.073$ (@~1.2)&0.24& 16\\
$J^\dagger$ &SMC&BL+WV&$-2.092\pm 0.116$&$15.495\pm 0.092$ (@~1.2)&0.33& 47\\
$\Ks^\dagger$ &SMC&BL&$-2.553\pm 0.444$&$16.924\pm 0.061$ (@~0.3)&0.32& 29\\
$\Ks^\dagger$ &SMC&WV&$-2.568\pm 0.247$&$15.041\pm 0.065$ (@~1.2)&0.22& 16\\
$\Ks^\dagger$ &SMC&BL+WV&$-2.113\pm 0.105$&$15.063\pm 0.082$ (@~1.2)&0.29& 45\\
$W_1(VI)$  &SMC&BL   &$-2.421\pm 0.479$&$16.832\pm 0.069$ (@~0.3)&0.26& 17\\
$W_1(VI)$  &SMC&   WV&$-3.003\pm 0.158$&$14.721\pm 0.040$ (@~1.2)&0.10& 10\\
$W_1(VI)$  &SMC&BL+WV&$-2.304\pm 0.107$&$14.789\pm 0.083$ (@~1.2)&0.23& 27\\
$W_2(VI)$  &SMC&BL   &$-2.430\pm 0.488$&$16.894\pm 0.070$ (@~1.2)&0.27& 17\\
$W_2(VI)$  &SMC&   WV&$-2.985\pm 0.153$&$14.811\pm 0.039$ (@~0.3)&0.10& 10\\
$W_2(VI)$  &SMC&BL+WV&$-2.277\pm 0.108$&$14.878\pm 0.085$ (@~1.2)&0.24& 27\\
\hline
$J$        &LMC&BL   &$-2.164\pm 0.240$&$17.131\pm 0.038$ (@~0.3)&0.25& 55\\
$J$        &LMC&WV   &$-2.337\pm 0.114$&$15.165\pm 0.030$ (@~1.2)&0.18& 82\\
$J$        &LMC&BL+WV&$-2.163\pm 0.044$&$15.194\pm 0.029$ (@~1.2)&0.21&137\\
$H$        &LMC&BL   &$-2.259\pm 0.248$&$16.857\pm 0.039$ (@~0.3)&0.26& 54\\
$H$        &LMC&WV   &$-2.406\pm 0.100$&$14.756\pm 0.027$ (@~1.2)&0.16& 82\\
$H$        &LMC&BL+WV&$-2.316\pm 0.043$&$14.772\pm 0.028$ (@~1.2)&0.20&136\\
$\Ks$      &LMC&BL   &$-1.992\pm 0.278$&$16.733\pm 0.040$ (@~0.3)&0.26& 47\\
$\Ks$      &LMC&WV   &$-2.503\pm 0.109$&$14.638\pm 0.029$ (@~1.2)&0.17& 82\\
$\Ks$      &LMC&BL+WV&$-2.278\pm 0.047$&$14.679\pm 0.029$ (@~1.2)&0.21&129\\
$W_1(VI)$&LMC&BL   &$-2.598\pm 0.094$&$16.597\pm 0.017$ (@~0.3)&0.10& 55\\
$W_1(VI)$&LMC&   WV&$-2.564\pm 0.073$&$14.333\pm 0.019$ (@~1.2)&0.11& 76\\
$W_1(VI)$&LMC&BL+WV&$-2.521\pm 0.022$&$14.339\pm 0.015$ (@~1.2)&0.11&131\\
$W_2(VI)$&LMC&BL   &$-2.572\pm 0.093$&$16.665\pm 0.016$ (@~0.3)&0.10& 55\\
$W_2(VI)$&LMC&   WV&$-2.551\pm 0.073$&$14.431\pm 0.019$ (@~1.2)&0.11& 76\\
$W_2(VI)$&LMC&BL+WV&$-2.486\pm 0.022$&$14.440\pm 0.015$ (@~1.2)&0.11&131\\
\hline
$J$&Globular&BL   &$-2.959\pm 0.313$&$-1.541\pm 0.041$ (@~0.3)&0.11&  7\\ 
$J$&Globular&WV   &$-2.204\pm 0.090$&$-3.543\pm 0.027$ (@~1.2)&0.16& 39\\ 
$J$&Globular&BL+WV&$-2.230\pm 0.053$&$-3.542\pm 0.024$ (@~1.2)&0.16& 46\\ 
$H$&Globular&BL   &$-2.335\pm 0.335$&$-1.847\pm 0.044$ (@~0.3)&0.12&  7\\ 
$H$&Globular&WV   &$-2.337\pm 0.086$&$-3.942\pm 0.025$ (@~1.2)&0.16& 39\\ 
$H$&Globular&BL+WV&$-2.344\pm 0.050$&$-3.944\pm 0.023$ (@~1.2)&0.15& 46\\ 
$\Ks$&Globular&BL   &$-2.294\pm 0.294$&$-1.864\pm 0.039$ (@~0.3)&0.10&  7\\ 
$\Ks$&Globular&WV   &$-2.442\pm 0.082$&$-3.997\pm 0.024$ (@~1.2)&0.15& 39\\ 
$\Ks$&Globular&BL+WV&$-2.408\pm 0.047$&$-4.004\pm 0.021$ (@~1.2)&0.14& 46\\ 
\hline
\end{tabular}
\end{center}
\end{minipage}
\end{table*}

Dispersions ($\sigma$) about the various PLRs are listed
in Table~\ref{tab:PLRs_T2C}.
For the LMC/SMC samples, these tend to be greater for the BLs than for the WVs.
Such differences though suggestive are of only marginal significance.
An extreme example is for the $\log P$ -$W_1~(VI)$
in the SMC where $\sigma$ is
0.26 for the BLs and 0.10~mag for the WVs. 
However, an F-test shows that this is not significant
at the 95~\% confidence level. In the case of the BL stars
in the SMC the C10 data alone shows a smaller scatter (0.26~mag) than
the IRSF data (0.37~mag) in $\Ks$, which probably indicates
a larger observational scatter
for the faintest stars in the latter sample.
It should be noted that in the case of the LMC we followed
\citet{Soszynski-2008b} and M09 in omitting a few bright BL stars  
as possible  binaries or blends, while no BL stars were omitted from the SMC
sample. Although no evidence of such binarity or blend has been reported
for the SMC BL stars, this may well affect the comparison of
the $\sigma$ values.

\subsection{Comments on the PLR slopes}
\label{sec:PLRslope}

\begin{figure*}
\begin{minipage}{170mm}
\begin{center}
\includegraphics[clip,width=0.85\hsize]{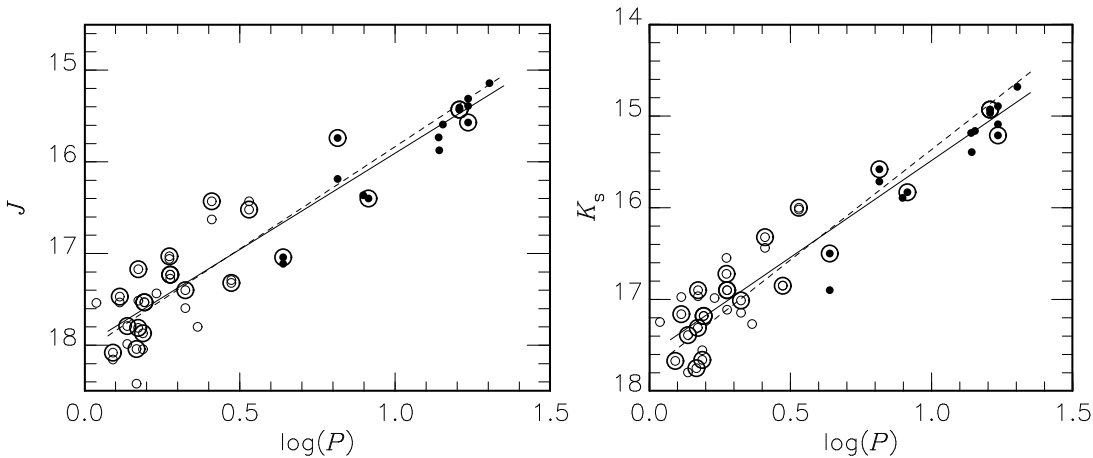}
\end{center}
\caption{
Period-luminosity relations in $J$ and $\Ks$ for
the combined IRSF and NTT (C10) dataset. 
BL and WV stars are indicated by open and filled circles
and those taken from \citet{Ciechanowska-2010} are encircled
in addition.
The solid lines indicate the least squares solutions, while 
the dashed lines show the relations used by C10 to derive an SMC
modulus.
\label{fig:combine_PLR}}
\end{minipage}
\end{figure*}

Fig.~\ref{fig:combine_PLR} shows PLRs in $J$ and $\Ks$ for the SMC
based on the combined sample of C10 and the present paper.
This consists of 16 measurements of BLs and 31 of WVs (29 at $\Ks$),
all of which were corrected for phase variation, for BLs and WVs.
C10 derived a distance modulus of 18.85~mag for the SMC 
by comparing their combined BL+WV sample with the PLRs in $J$ and $\Ks$
derived for globular cluster T2Cs by M06. 
In Fig.~\ref{fig:combine_PLR} these relations are shown 
as dashed
lines for the distance derived by C10.
The solid lines are least square fits to the data. Apparently
the dashed 
lines with the cluster slopes do not fit the data.
This is particularly noticeable in $\Ks$
where the best fit with a line of cluster slope is too bright
by 0.21~mag at $\log P = 1.3$ and too faint by 0.09~mag at $\log P = 0.3$.
This discrepancy, although present,
is not as clear from C10 data alone since they had only two long period
WV stars. 

The slopes listed in Table~\ref{tab:PLRs_T2C} show that, in fact, there
is a general trend for the PLR slopes to become more negative as one goes
from the SMC to the LMC and to the globular clusters. 
The BL+WV samples have the slopes, in the $\Ks$-band, of 
$-2.113\pm 0.125$, $-2.278 \pm 0.047$ and $-2.408 \pm 0.047$,
and in $W_1(VI)$ for the SMC and LMC 
the slopes are $-2.304 \pm 0.107$ and $-2.521 \pm 0.022$.
It should be noted, on the other hand, that the slopes of either the BLs
or of the WVs taken by themselves are very uncertain
due to the small period range and the relatively small number
of stars in each group.
The difference in slope between systems may be caused by metallicity
and/or age effects
although these parameters of the T2Cs are unknown for the Magellanic Clouds.
Evidently an SMC distance, and to a lesser extent
an LMC distance, based on the cluster data will vary according to the period 
at which the comparison is made.

\section{The distance difference between the Magellanic Clouds}
\label{sec:delmu}

\subsection{Possible effects of Magellanic Cloud structure}
\label{sec:SMCdepth}

\begin{figure*}
\begin{minipage}{170mm}
\begin{center}
\includegraphics[clip,width=0.70\hsize]{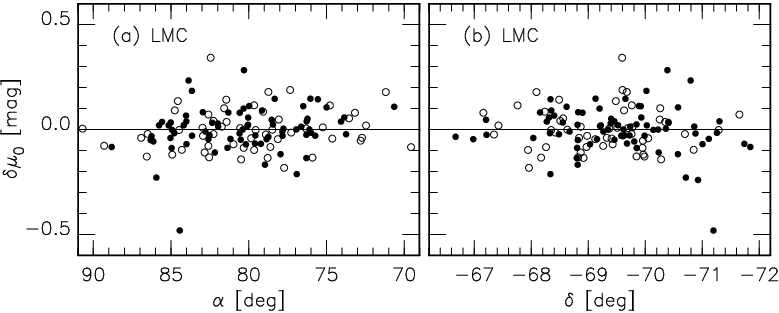}
\includegraphics[clip,width=0.70\hsize]{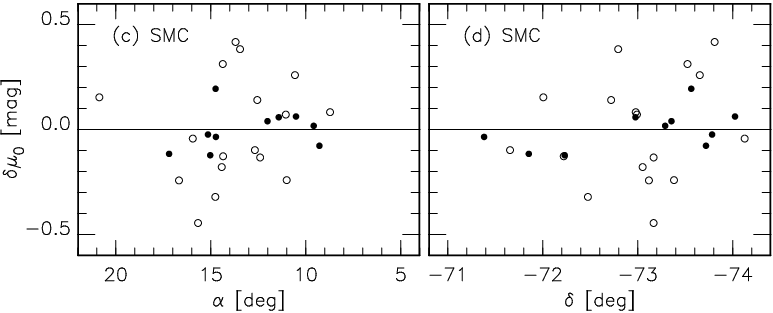}
\end{center}
\caption{The scatter of the distance moduli of individual T2Cs
in the LMC, panels (a) and (b), and in the SMC, (c) and (d)
plotted against Right Ascension and Declination.
BL and WV stars are indicated by open and filled circles.
The horizontal lines indicate the barycentres
which are determined by the PLRs.
A positive $\delta\mu_0$ means that the object is more distant
than the centre.
\label{fig:DMscatter2}}
\end{minipage}
\end{figure*}

The Magellanic Clouds are close enough to us that their three
dimensional structure could affect the relative mean distance
of different groups of stars   
These structures have been studied using
several tracers: e.g.~classical Cepheids (e.g.~\citealt{Caldwell-1986};
\citealt{Groenewegen-2000}),
red variables (Lah, Kiss \& Bedding, 2005) 
and red-clump giants \citep{Subramanian-2009}. 
These authors discussed the internal structures of the Clouds,
especially a large line-of-sight depth and substructures of the SMC.
Fig.~\ref{fig:DMscatter2}
plots the deviations, $\delmu$, from the $\log P$-$W_1 (VI)$ relation
against Right Ascension and Declination for both Clouds.
We used the PLRs derived for the BLs and WVs separately 
to estimate their $\delmu$ values. 
These plots are similar to those for classical Cepheids 
(Fig.~\ref{fig:DMscatter}). In particular the spread is much wider
for the SMC than for the LMC due to its large depth.
Other major features known from the classical Cepheids are also seen in
the T2Cs; e.g. the north-east part of the SMC is closer
than the main body. The apparent larger scatter for the
BLs compared with the WVs in the SMC may not be
significant (see Section~\ref{sec:SMC_T2C_PLR}).
The zero points of the PLRs correspond to the barycentres
of the whole sample, which are indicated by 
the horizontal lines in Fig.~\ref{fig:DMscatter2} and \ref{fig:DMscatter}.
The distances discussed in the following refer to these barycentres.
 
\subsection{The distance difference from the PLRs of Cepheids}
\label{sec:PLcomparison}
In view of the results of Section~\ref{sec:PLRslope} we have derived
the difference
in moduli of the two Clouds independently for the BLs and WVs.
This minimizes any effect on the results of the choice of slopes for
PLRs. In this section we also compare these results with those
for fundamental mode, classical Cepheids, 
using the results of Appendix~\ref{sec:classicalCep}.
For both types of Cepheids, 
the samples were collected from
the same datasets: the OGLE-III variability search in the optical and
the IRSF survey in the near-IR.

Table~\ref{tab:deltamu} compares the 
$\delmu$ values, in $\Ks$ and $W(VI)$,
from the classical Cepheids and the T2Cs.
The $\delmu$ are estimated individually for the groups with
different period ranges:
three groups with the thresholds of 0.4 and 1.0
in $\log P$ for classical Cepheids,
and BL ($\log P<0.6$) and WV ($\log P > 0.6$) for T2Cs.
We here consider the $\Ks$-band PLR for the SMC obtained 
for the combined datasets of C10 and ours, but
the results are not affected if only our photometry is used.
In the use of the $\Ks$-band PLR, we make a corrections for extinctions
of $E_{B-V}=0.074$ and 0.054 for the LMC and SMC respectively
\citep{Caldwell-1985} according to the extinction law of \citet{Cardelli-1989}.

\begin{table}
\begin{minipage}{0.95\hsize}
\caption{
The SMC-LMC modulus difference ($\delmu$)
derived from the PLRs of T2Cs and classical Cepheids.
No corrections for metallicity effects have been applied.
\label{tab:deltamu}}
\begin{center}
\begin{tabular}{cccc}
\hline
Type & \multicolumn{3}{c}{$\delmu$} \\
(or $\log P$ range) &from $\Ks$&from $W_1(VI)$&from $W_2(VI)$\\
\hline
\multicolumn{4}{c}{Type II Cepheids} \\
BL&$0.19\pm 0.07$&$0.24\pm 0.07$&$0.23\pm 0.07$\\ 
WV&$0.40\pm 0.07$&$0.39\pm 0.05$&$0.38\pm 0.04$\\ 
BL+WV&$0.38\pm 0.11$&$0.45\pm 0.08$&$0.44\pm 0.08$\\ 
\multicolumn{4}{c}{Classical Cepheids} \\
0.0--0.4&$0.530\pm 0.012$&$0.543\pm 0.009$&$0.537\pm 0.010$\\
0.4--1.0&$0.484\pm 0.008$&$0.482\pm 0.008$&$0.476\pm 0.008$\\
1.0--1.7&$0.478\pm 0.025$&$0.486\pm 0.025$&$0.479\pm 0.024$\\
0.4--1.7&$0.495\pm 0.011$&$0.484\pm 0.010$&$0.480\pm 0.009$\\
\hline
\end{tabular}
\end{center}
\end{minipage}
\end{table}

\subsubsection{$\delmu$ from classical Cepheids}

For the classical Cepheids with $\log P > 0.4$ the results agree well
with previous estimates (e.g.~\citealt{Groenewegen-2000}).
A larger $\delmu$ value is indicated at the shorter periods.
In the SMC, the PLR slopes at these short periods are different
from the longer period ones. This is not the case in the LMC.
This suggests that they are less suitable for measuring distance differences
at least in the LMC/SMC metallicity range
(see also Appendix~\ref{sec:classicalCep}).

Metallicity effects on the PLRs of classical Cepheids
have been the subject of many publications.
On the observational side, two methods are frequently used:
(i)~comparing Cepheids located in a galaxy
at different distances from its centre and thus characterized
by different abundances 
(e.g.~\citealt{Freedman-1990}; \citealt{Macri-2006};
\citealt{Scowcroft-2009}), and (ii)~comparing Cepheid-based distances
with those obtained with
an independent distance indicator, most often the tip of 
Red Giant Branch (RGB)
(e.g.~\citealt{Sakai-2004}; \citealt{Bono-2010}).
Recently, \citet{Bono-2010} investigated the metallicity
effects on the PLRs of classical Cepheids 
from both the observational and theoretical sides and
found that these effects were small in
$W(VI)$ for a broad range of metallicities.
Taking into account estimates in the literature, we consider here
a range in the metallicity effect of 0 to $-0.3$~mag~dex$^{-1}$,
for the classical Cepheids with $\log P > 0.4$.
The metallicity difference between the Clouds,
$\sim 0.4$~dex \citep{Romaniello-2008}, leads to the correction 
of 0--0.1~mag. We adopt a mean value, i.e.~$0.05\pm 0.05$~mag,
which leads to a metallicity corrected value of
$\delmu^{\rm cor}=0.43\pm 0.05$~mag.
The error is dominated by the uncertainty of the metallicity correction.

\begin{table*}
\begin{minipage}{0.95\hsize}
\caption{
The SMC-LMC modulus difference, $\delmu$,
derived in various ways.
The values corrected for metallicity and/or other population effects,
$\delmu^{\rm cor}$, are also listed. Their errors include the uncertainty of
this correction. References are indicated in the last column.
\label{tab:deltamu2}}
\begin{center}
\begin{tabular}{ccccc}
\hline
Method & $\delmu$ & $\delmu^{\rm cor}$ & Notes & Ref. \\
\hline
Type II Cepheids (WV)  & $0.40\pm 0.07$ & --- & (1) & This work \\
                   & $0.39\pm 0.05$ & --- & (2) & This work \\
Classical Cepheids & $0.48\pm 0.01$ & $0.43\pm 0.05$ & (1,3,4) & This work \\
                   & $0.48\pm 0.01$ & $0.43\pm 0.05$ & (2,3,4) & This work \\
                   & $0.48\pm 0.04$ & $0.43\pm 0.06$ & (1,3) & \citet{Groenewegen-2000} \\
                   & $0.51\pm 0.02$ & $0.46\pm 0.05$ & (2,3) & \citet{Groenewegen-2000} \\
                   & $0.41\pm 0.20$ & $0.39\pm 0.20$ & (1) & \citet{Bono-2010} \\
                   & $0.45\pm 0.12$ & $0.44\pm 0.12$ & (2) & \citet{Bono-2010} \\
RR~Lyrae variables & $0.327\pm 0.002$ & $0.363\pm 0.04$ & (1,5) & \citet{Szewczyk-2009} \\
Red clump giants  & $0.47\pm 0.03$ & 0.43    &  (1,6) & \citet{Pietrzynski-2003} \\
                  & $0.38\pm 0.02$ & $0.45\pm 0.10$ &  (2) & \citet{Pietrzynski-2003} \\
Tip of RGB & --- & $0.40\pm 0.12$ & (2,7) & \citet{Sakai-2004}  \\
Pulsating M giants & $0.41\pm 0.02$ & --- & (1,8) & \citet{Tabur-2010} \\
Eclipsing binaries & --- & $0.5\pm0.15$ & (9) & ---       \\
\hline
\end{tabular}
\end{center}
Notes: (1)~Based on near-IR photometry, $\Ks$ or $K$.
(2)~Based on optical photometry, $I$ or $W(VI)$.
(3)~Our adopted metallicity correction, 0--0.1~mag,
was used to derive $\delmu^{\rm cor}$.
(4)~Only those with $0.4 < \log P < 1.0$ were used.
(5)~Taking the means of the values based on different calibrations
discussed in \citet{Szewczyk-2009}.
(6) The population correction suggested by \citet{Salaris-2002} was used.
(7)~ A metallicity correction was derived using the colour of the RGB.
(8)~\citet{Tabur-2010} concluded that the metallicity effect is negligible.
(9)~This value is a rough estimate (see text).
\end{minipage}
\end{table*}

\subsubsection{$\delmu$ from type II Cepheids}

No theoretical predications
seem to have been made for metallicity effects for T2Cs.
However, the results for T2Cs in globular clusters (M06), which are heavily
weighted to WV stars, suggests that at $\Ks$ any such effects
are small. The present results support this conclusion since
the $\delmu$ values for the WV stars in Table 3 are 
in satisfactory agreement with the metallicity corrected value
for the longer period classical Cepheids.

On the other hand, the $\delmu$ value from the BL stars is smaller than
for the WVs or classical Cepheids.
This is a result at the $2~\sigma$ level and is present both in $\Ks$ and $VI$.
This points to a difference in mean properties of BLs
in the LMC and SMC. Mean metallicity or age differences are
the most likely candidates.

\subsection{Comparison with other methods}

Various other estimates of the difference in the distance moduli of
the Clouds, $\delmu$, are listed in Table~\ref{tab:deltamu2}.
This table also shows the results for WV stars, and 
the longer period classical Cepheids. 

Szewczyk~et~al.~(\citeyear{Szewczyk-2008}, \citeyear{Szewczyk-2009})
used near-IR PLRs of RR~Lyraes
to derive the distance moduli of the Clouds.
They adopted mean metallicities for the LMC and SMC RR~Lyraes of
$-1.48$ and $-1.7$~dex. This leads to
changes in $\delmu$ of $0.02$--$0.05$~mag
depending on which metallicity dependent calibration of the PLR
is adopted. These corrections bring the difference into better
agreement with the WVs and classical Cepheids.
Interestingly, the metallicity correction for RR~Lyraes has the same sign as
that suggested for BLs above.

Hilditch, Howarth \& Harries (2005) 
obtained an SMC distance modulus of $18.91\pm 0.1$~mag
from eclipsing binaries,
while a larger value, $19.11 \pm 0.03$~mag, was obtained by
\citet{North-2010}. 
The difference in these values may come partly from the internal structure of
the SMC, but other factors seem to be involved.
One of these is the use of different surface brightness-colour relations.
These tend to be especially uncertain for the early type binaries
which have generally been used.
The method is also sensitive to reddening estimates.
The available distance estimates for the LMC from early type binaries
show a rather large scatter (see e.g.~\citealt{Nelson-2000}).
\citet{Pietrzynski-2009} derived an LMC distance modulus of
$18.50\pm 0.06$ based on a binary system composed of two late-type giants.
In Table 4, $\delmu =\sim 0.5\pm 0.15$~mag is adopted. 

For red clump stars, Pietrzy\'{n}ski, Gieren \& Udalski (2003) 
obtained $\delmu = 0.47 \pm 0.03$~mag from $K$ magnitudes. 
A population correction by \citet{Salaris-2002} leads to
0.43~mag.
\citet{Tabur-2010} compared the PLR sequences of pulsating M-type giants,
i.e.~semi-regulars and Miras, in the Magellanic Clouds and those in
the solar neighbourhood, finding $\delmu=0.41\pm 0.02$~mag.
They concluded that the metallicity effect for their PLRs is
negligible.

In summary, the mean uncorrected $\delmu$ for the WVs is $0.39\pm 0.05$~mag.
This agrees well with the other results in Table~\ref{tab:deltamu2}.
These, corrected for metallicity and age effects where necessary, range from 0.36 to 0.46~mag 
(while eclipsing variables giving $\sim 0.5$~mag).
Thus any effect due to population differences on the WVs appear to be small.
On the other hand, the smaller $\delmu$ for the BLs, together with
the result shown in Fig.~\ref{fig:combine_PLR},
suggests that the population difference between the two Clouds is
affecting the mean luminosities of these stars.

\section{The periods and colours of type II Cepheids in different systems}
\label{sec:evolution}

\subsection{Period distribution of type II Cepheids}

\begin{figure}
\begin{center}
\begin{minipage}{85mm}
\begin{center}
\includegraphics[clip,width=0.80\hsize]{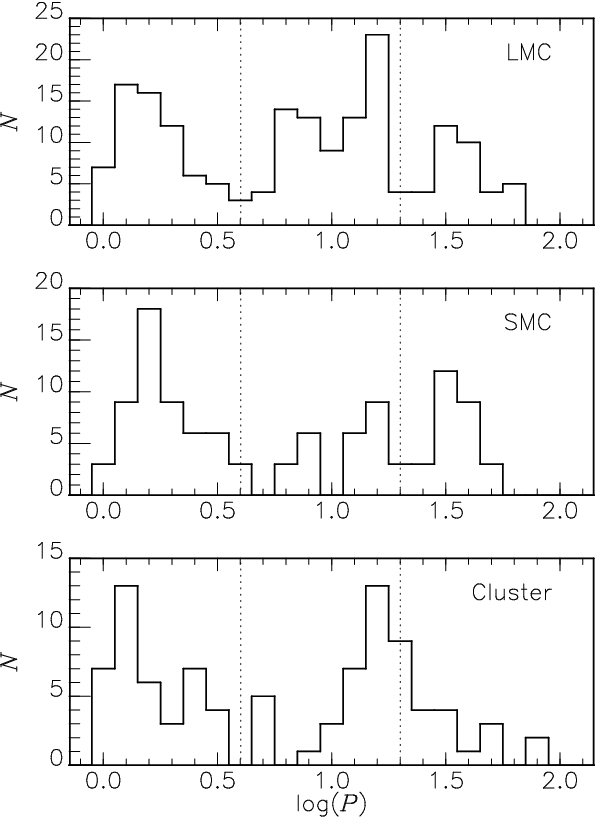}
\end{center}
\caption{
Histograms of periods for the T2Cs in the LMC (top),
SMC (middle) and globular clusters (bottom). The pW stars are not included.
Vertical lines show the thresholds, 4 and 20~d,
used to divide BL, WV and RV variables.
\label{fig:Phist3}}
\end{minipage}
\end{center}
\end{figure}

Fig.~\ref{fig:Phist3} compares the period distributions of
the T2Cs in the Magellanic Clouds
and in globular clusters. The pW stars are not included
in this comparison.
The sample for globular clusters has been slightly changed from
that in M09
to include the T2Cs in
NGC~6388 and 6441 (table~7 in \citealt{Pritzl-2003}).

A separation between WV and RV stars is evident
for the LMC and SMC samples, while the distribution of
the cluster sample suggests that the stars in the RV period
range are just the tail  
of the WV star distribution. This together with the
fact that cluster variables in the RV period range lie on
an extension of the cluster T2C PLR indicates a fundamental
difference from those in the Clouds which lie about PLRs
fitted to shorter period T2Cs.
There is a suggestion that the relative frequency of
variables with log P$\sim 0.8$ is greater in the LMC than
the clusters.

\begin{figure}
\begin{center}
\begin{minipage}{85mm}
\begin{center}
\includegraphics[clip,width=0.85\hsize]{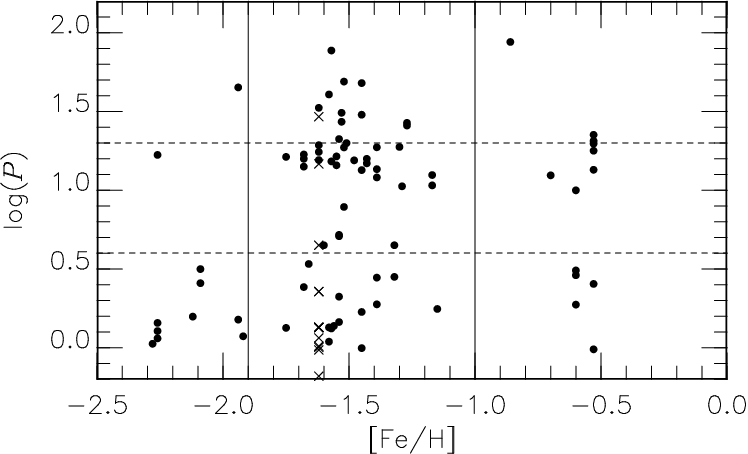}
\end{center}
\caption{
Periods and metallicities of the T2Cs in globular clusters.
The cross symbols indicate those in $\omega$~Cen plotted at
its mean metallicity.
The horizontal lines show the 
divisions between the BL, WV and RV stars, and
the vertical lines the adopted divisions
by metallicity (see text).
\label{fig:FeP}}
\end{minipage}
\end{center}
\end{figure}

There is a difference in
the relative frequencies of 
WV and BL stars between the LMC and the SMC.
The ratio of the two types, $N_{\rm WV}/N_{\rm BL}$, is $1.25\pm 0.2$
for the LMC and $0.6\pm 0.2$ for the SMC, where the Poisson noise
is considered as the uncertainty.
Although the LMC ratio changes
if we exclude the conspicuous group of the LMC WV stars with
$\log P \sim 0.8$, it would still be larger than the ratio for the SMC.
To test for a metallicity effect in this ratio,
the globular cluster sample was divided into three groups: 11 objects with
${\rm [Fe/H]}<-1.9$,
53 with $-1.9<{\rm [Fe/H]}<-1.0$ and 13 with ${\rm [Fe/H]}>-1$~dex.
Fig.~\ref{fig:FeP} plots the periods of the globular T2Cs against their
metallicities (taken from \citealt{Harris-1996}).
We do not include in the discussion the objects in $\omega$~Cen,
indicated by the cross
symbols in Fig.~\ref{fig:FeP}, because this cluster has a significant spread
in [Fe/H].
Adopting $P=4, 20$~d as the dividing lines of the T2C types,
$N_{\rm WV}/N_{\rm BL}$ are $0.1 \pm 0.1, 1.9\pm 0.6$ and $1.0\pm 0.6$
for the three groups. The sample in the most metal-rich range is
dominated by NGC~6388 and 6441, the two metal-rich clusters with
a peculiar population of RR~Lyrae stars \citep{Pritzl-2003}.
Comparing the WV/BL ratios between
the other two metallicity ranges suggests that a metal-poor
stellar system has a lower WV/BL ratio. This is qualitatively
consistent with above results if the SMC T2Cs are lower
in metallicity than the LMC ones.
Further discussion will be possible when metallicities
have been determined for T2Cs in the Magellanic Clouds.

It is interesting that
the ratio of long-period to short-period stars
is also greater in the LMC than in the SMC for the classical Cepheids 
(\citealt{Soszynski-2010a}; also see
Appendix \ref{sec:classicalCep}).
This has generally been attributed to a metallicity effect
(Becker, Iben \& Tuggle, 1977). 
Although not included in Fig.~\ref{fig:Phist3},
the SMC pW stars have longer periods
than their counterparts in the LMC as pointed out in S10
and as can be seen from Fig.~\ref{fig:compPK}.
The reason for this is not clear.
No counterpart to pW stars have been identified in globular clusters.

\subsection{Colours of type II Cepheids}

\begin{figure*}
\begin{center}
\begin{tabular}{cc}
\begin{minipage}{70mm}
\begin{center}
\includegraphics[clip,width=0.85\hsize]{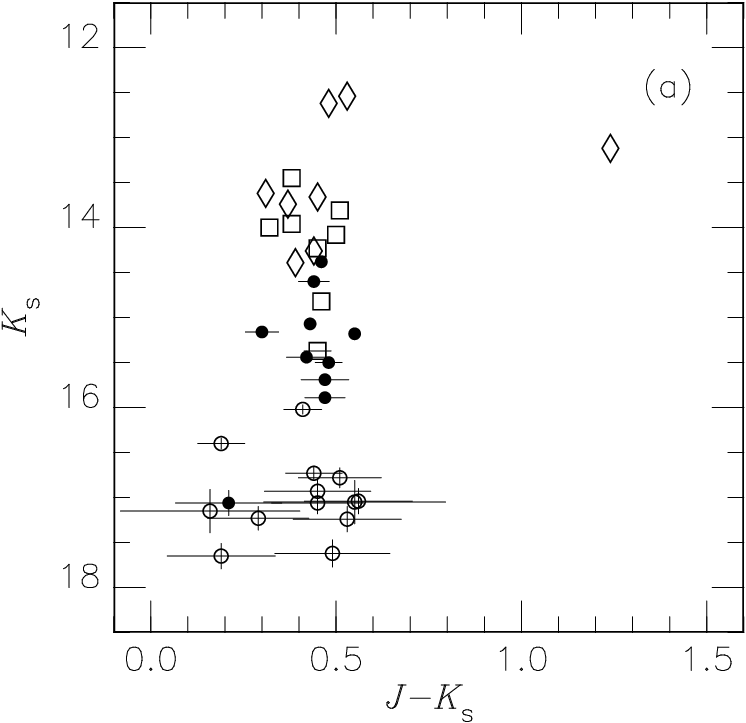}
\end{center}
\end{minipage}
\begin{minipage}{70mm}
\begin{center}
\includegraphics[clip,width=0.85\hsize]{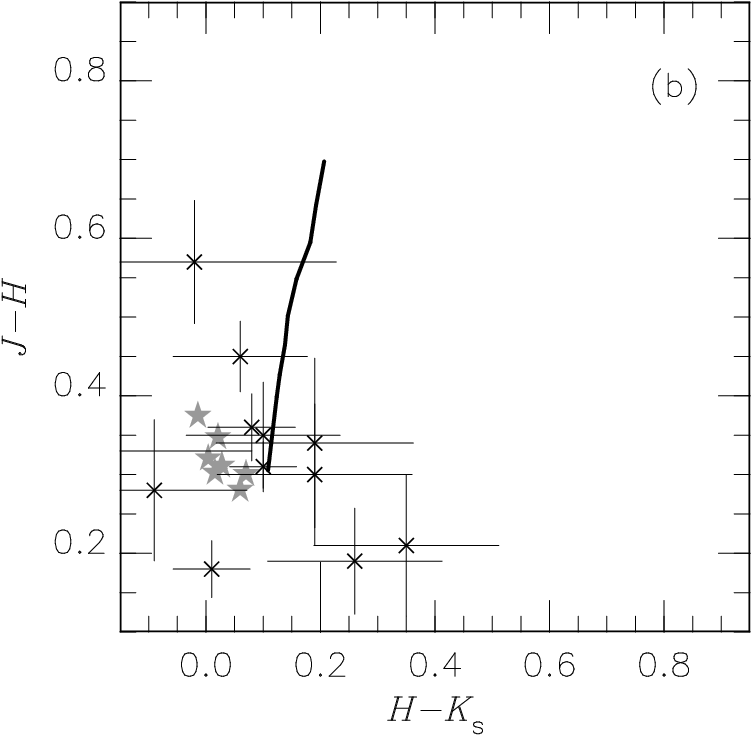}
\end{center}
\end{minipage}
\vspace{2mm}
\\
\begin{minipage}{70mm}
\begin{center}
\includegraphics[clip,width=0.85\hsize]{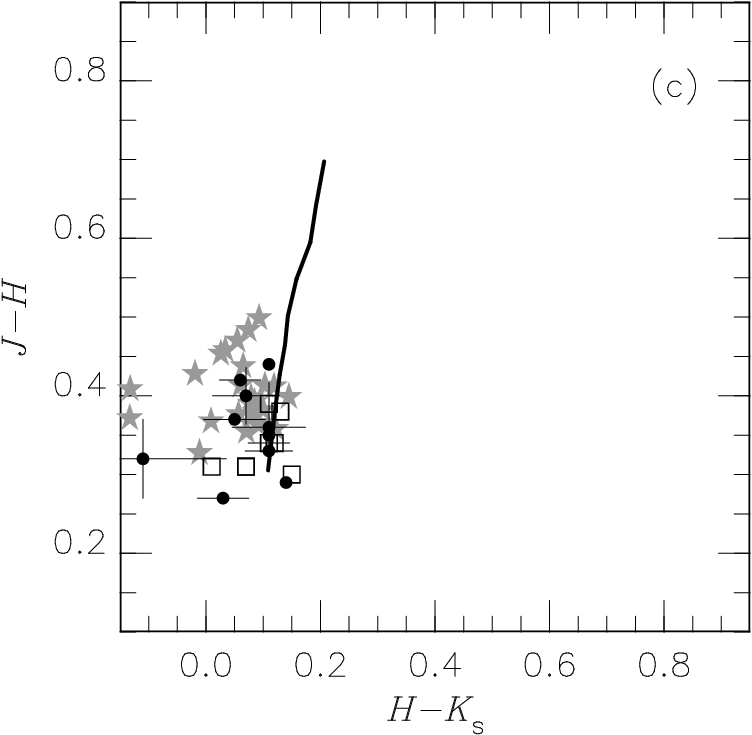}
\end{center}
\end{minipage}
\begin{minipage}{70mm}
\begin{center}
\includegraphics[clip,width=0.85\hsize]{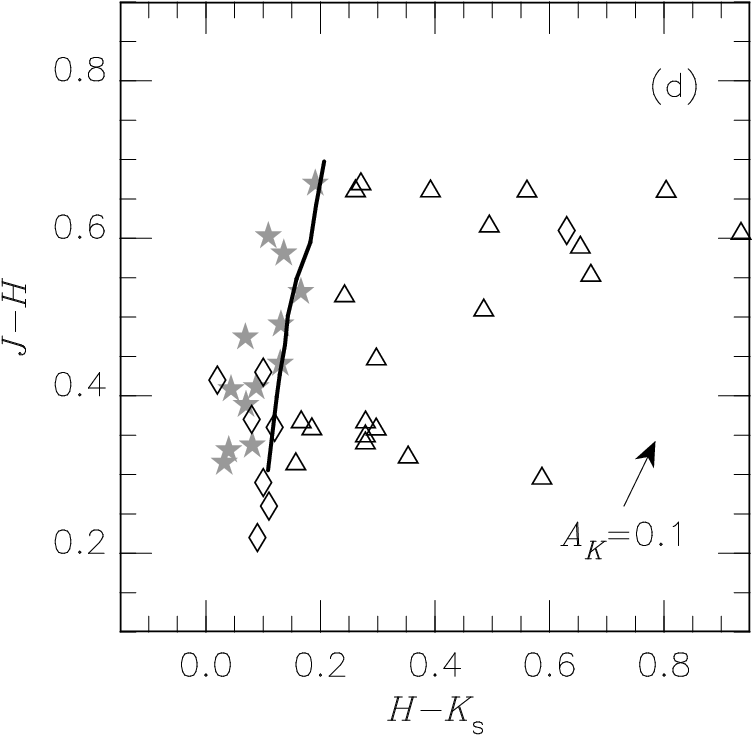}
\end{center}
\end{minipage}
\end{tabular}
\caption{
A colour-magnitude diagram and colour-colour diagrams for T2Cs.
Panel (a) includes all types SMC T2Cs,
while panels (b), (c) and (d) are for
BL, WV/pW and RV stars.
Symbols for the SMC T2Cs are the same as in Fig.~\ref{fig:PLR}.
Grey star symbols indicate the T2Cs in globular clusters (M06).
In panel (d), the triangles are for Galactic RV stars
from \citet{LloydEvans-1985}.
Error bars are drawn for the SMC T2Cs
only if an uncertainty exceeds the size of the symbol.
The thick curve in the colour-colour diagrams is the locus of
local giants from \citet{Bessell-1988}.
\label{fig:CMDCCD}}
\end{center}
\end{figure*}

Fig.~\ref{fig:CMDCCD} shows colour-magnitude and
colour-colour diagrams
for BL, WV, pW and RV variables in the SMC.
T2Cs in globular clusters are plotted by grey star symbols
in the colour-colour diagrams for comparison.
In panel (d), the triangles indicate Galactic RV stars
from \citet{LloydEvans-1985}.
Their magnitudes were converted from the SAAO system
into that of the IRSF using transformations in
\citet{Carpenter-2001} and \citet{Kato-2007}.
For the cluster variables, the dereddened colours have been adjusted for the
mean SMC reddening ($E_{B-V}=0.054$~mag, \citealt{Caldwell-1985}).
The colours of the Galactic RV stars have not been corrected
for their reddenings.
The curves in the two-colour plots indicate the location of normal
giants (G0III--K5III) taken from \citet{Bessell-1988} whose colours are also
transformed into the IRSF system after adding the SMC reddening.
The distribution of the SMC objects in these plots is similar
to that of the LMC objects (fig.~3 in M09).

\begin{figure}
\begin{minipage}{85mm}
\begin{center}
\includegraphics[clip,width=0.85\hsize]{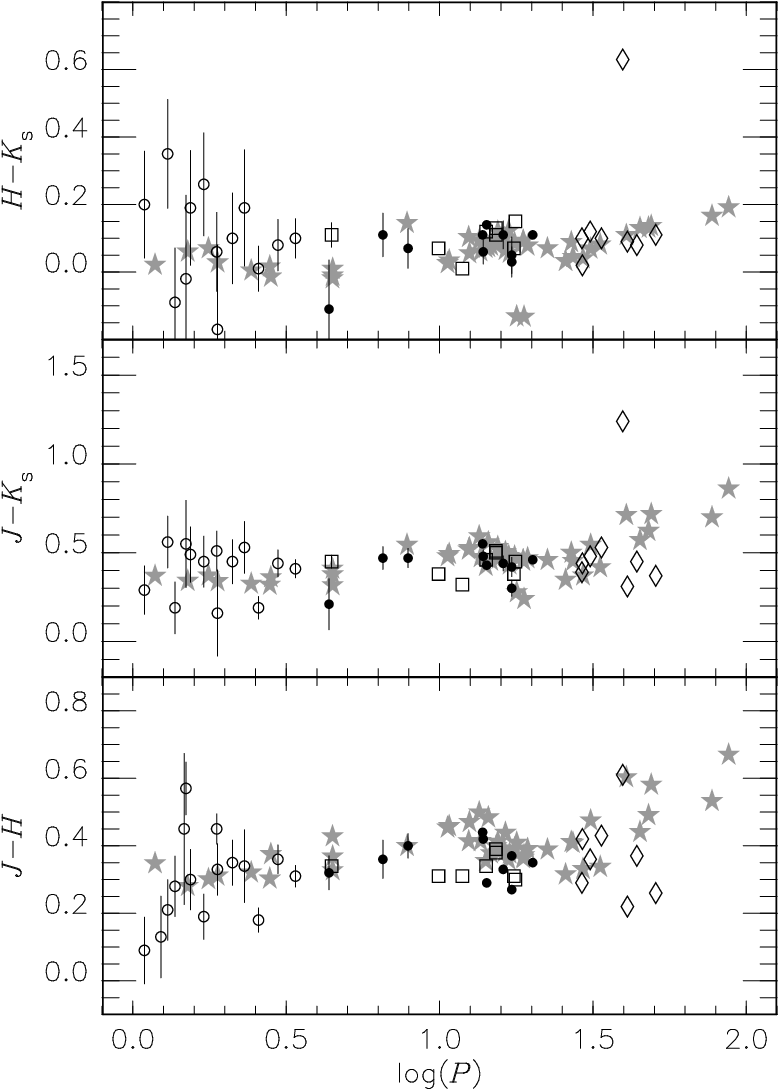}
\end{center}
\caption{
Period-colour relations for T2Cs.
Symbols are the same as in Fig.~\ref{fig:CMDCCD}.
Error bars are indicated for the SMC T2Cs
only if the uncertainties exceed 0.03~mag.
\label{fig:PCR}}
\end{minipage}
\end{figure}

Infrared colour/$\log P$ relations are plotted in Fig.~\ref{fig:PCR}.
No correction for interstellar reddening was applied to
the SMC objects.
The dereddened
magnitudes of the globular cluster objects have been reddened by
an amount corresponding to the SMC reddening as in Fig.~\ref{fig:CMDCCD}, 
so that the two samples are directly comparable.
The WV stars in globular clusters, $1.0\leq \log P \leq 1.3$,
have a trend for $(J-H)$ and $(J-\Ks)$ to become bluer with
increasing period, which is not clear for the SMC WVs.
Two pW stars in the SMC, \#1 and \#11, are separated from the trend
of globular WV stars.
The lower photometric accuracies for BL stars prevent us from any comparison
with the clusters.

RV stars show a wider colour distribution and appear to be
heterogeneous in nature (see section~5.2 in M09).
An SMC RV star, \#18, is distinctly red in $J-\Ks$ and $H-\Ks$
but not in $J-H$.
A few RV stars are similar to the globular objects 
in $H-\Ks$ but bluer in $J-H$ at a given period.
These features of the SMC T2Cs are similar to those in the LMC (M09). 
On the other hand, the two brightest stars, RV stars \#7 and \#29,
are not exceptionally red in any near IR colour. 
This is different from the situation in the LMC where the three brightest
RV stars have an excess in $\Ks$ (M09)
as does the SMC RV star \#18. In the $\log P$-$\Ks$ plot
(Fig.~\ref{fig:compPK}) \#7 and \#29 are almost
as bright as classical Cepheids with similar periods
(see Fig.~\ref{fig:compPK}). 
S10 notes that these two stars are in binary, eclipsing or ellipsoidal, systems.

\section{Conclusions}
\label{sec:Conclusion}

In this paper PLRs are derived for T2Cs in the SMC.
Evidence is presented that T2C PLR slopes 
differ in three different systems
(globular clusters, LMC, SMC) when BL and WV stars
are used in combined solutions.
Treating the BL and WV stars separately, it was found that the
difference in distance moduli between the SMC and LMC
derived from WVs alone agrees closely with that obtained
from other distance indicators. This implies that
the absolute magnitudes of the WVs are, within the uncertainties,
free of metallicity effects, which is in agreement with the results
derived from globular clusters. On the other hand the SMC-LMC
difference is smaller for the BLs and suggests that their
absolute magnitudes are not immune from population effects. 
The relative frequencies of WV to BL stars also varies from
system to system.
\section*{Acknowledgments}
This work is based on the result of the OGLE project.
The OGLE has received funding from the European Research Council
under the European Community's Seventh Framework Programme
(FP7/2007-2013)/ERC grant agreement no.~246678.
We thank the referee (G.~Bono) for his useful comments.
NM acknowledges the support by the Grant-in-Aid for Research
Activity Start-up (No.~22840008) from the Japan Society for
the Promotion of Science (JSPS).
The work of MWF is supported by
South African National Research Foundation.

\appendix

\section{The PLRs of classical Cepheids}
\label{sec:classicalCep}
For comparison with the T2Cs, PLRs have been derived for the
classical Cepheids in both Clouds found in the OGLE III survey
(\citealt{Soszynski-2008a}; \citealt{Soszynski-2010a}).
Choosing fundamental mode pulsators only there are
1849 stars in the LMC and 2626 in the SMC.
Of these, there are matches 
within $0.5^{\prime\prime}$, for 1797 and 2436 stars
in the IRSF catalogue (Table~\ref{tab:catalogueI}).
Among these, 140 and 187 stars were detected in the IRSF survey twice
because they are located
in overlapping fields of the IRSF survey.
These duplicates show that
the phase-correction procedure which gives satisfactory results for
the T2Cs does not work
well for the classical Cepheids. This is probably because of
systematically different light curves in $I$ and in $\JHK$
for the latter.
We thus use the single-epoch IRSF magnitudes to discuss the near-IR PLRs.
This has a negligible effect on the final results because
of the large number of stars involved.

\begin{table*}
\begin{minipage}{0.95\hsize}
\caption{
The catalogue of OGLE-III classical Cepheids, pulsating in the fundamental mode, with IRSF counterparts.
The periods and corresponding IRSF measurements are listed as in Table~\ref{tab:catalogue}.
Shifts for the phase corrections ($\delta_{\phi}$) obtained from the $I$-band light curves are also listed, but not used in our plots and fits.
This is the first 10 lines of the full catalogue which will be available in the online version of the article (see Supporting Information).
Note that both the LMC and SMC objects are included in a single catalogue, but each OGLE-ID indicates one of the Clouds.
\label{tab:catalogueI}}
\begin{center}
\begin{tabular}{cccccccccccr}
\hline
OGLE-ID & $\log P$ & \multicolumn{9}{c}{IRSF counterpart} & \multicolumn{1}{c}{$\delta _{\phi}$} \\
 & & IRSF-Field & MJD(obs) & Phase & $J$ & $E_J$ & $H$ & $E_H$ & $\Ks$ & $E_\Ks$ & \\
\hline
OGLE-LMC-CEP-0028 & $ 0.10139$ & LMC0443-6940F & 53036.771 & 0.844 & 16.34 & 0.03 & 16.07 & 0.03 & 15.95 & 0.08 & $-0.169$ \\
OGLE-LMC-CEP-0033 & $ 0.85617$ & LMC0444-6920C & 53036.877 & 0.068 & 13.72 & 0.02 & 13.45 & 0.01 & 13.38 & 0.02 & $ 0.177$ \\
OGLE-LMC-CEP-0034 & $ 1.05133$ & LMC0442-7040E & 52693.855 & 0.645 & 13.23 & 0.02 & 12.84 & 0.01 & 12.71 & 0.02 & $-0.165$ \\
OGLE-LMC-CEP-0040 & $ 0.71308$ & LMC0443-6940C & 53034.803 & 0.132 & 13.99 & 0.01 & 13.74 & 0.01 & 13.70 & 0.02 & $ 0.191$ \\
OGLE-LMC-CEP-0042 & $ 0.41112$ & LMC0442-7020H & 52692.889 & 0.978 & 14.99 & 0.01 & 14.87 & 0.02 & 14.80 & 0.02 & $ 0.262$ \\
OGLE-LMC-CEP-0042 & $ 0.41112$ & LMC0443-7000B & 53039.905 & 0.635 & 15.26 & 0.01 & 14.96 & 0.02 & 14.90 & 0.03 & $-0.174$ \\
OGLE-LMC-CEP-0046 & $ 0.94663$ & LMC0444-6920H & 53498.715 & 0.522 & 13.47 & 0.02 & 13.08 & 0.02 & 12.99 & 0.01 & $-0.117$ \\
OGLE-LMC-CEP-0048 & $ 0.74418$ & LMC0443-6940E & 53034.855 & 0.166 & 13.85 & 0.01 & 13.57 & 0.01 & 13.52 & 0.02 & $ 0.145$ \\
OGLE-LMC-CEP-0049 & $ 0.58536$ & LMC0442-7040D & 52692.971 & 0.257 & 14.67 & 0.02 & 14.33 & 0.01 & 14.24 & 0.03 & $ 0.028$ \\
OGLE-LMC-CEP-0050 & $ 0.89463$ & LMC0443-6940H & 53037.841 & 0.805 & 13.81 & 0.02 & 13.42 & 0.02 & 13.34 & 0.03 & $-0.164$ \\
\hline
\end{tabular}
\end{center}
\end{minipage}
\end{table*}

\begin{figure*}
\begin{minipage}{170mm}
\begin{center}
\includegraphics[clip,width=0.7\hsize]{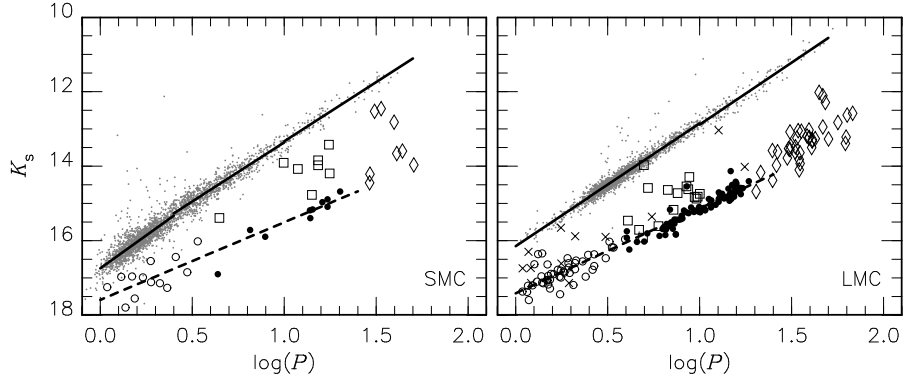}
\end{center}
\caption{
The PLRs at $\Ks$ for classical Cepheids in the
SMC(left) and LMC  (right) (grey dots and filled lines,
see Appendix~\ref{sec:classicalCep}).
Also shown for comparison are the
relations for the T2Cs, indicated by dashed lines.
Symbols for the T2Cs are the same as in Fig.~\ref{fig:PLR}
except the crosses in the LMC sample which indicate stars
not used in the least squares fits (see M09).
RVs and pWs are shown as open circles and squares.
\label{fig:compPK}}
\end{minipage}
\end{figure*}

Fig.~\ref{fig:compPK} shows the $\Ks$-band PLRs
of the classical Cepheids in the both Clouds.
Those for the T2Cs are also plotted.
Classical Cepheids are brighter than T2Cs by 1.0--2.5~mag depending
on the periods.
The pW/RV stars are located between the PLRs of
two types of Cepheids.

The classical Cepheids in each galaxy were divided into
three groups according to their periods using divisions
at $0.4$ and $1.0$ in $\log P$.
For each group and period range,
a least-square fit was obtained with a $3\sigma$ clipping applied.
(Table~\ref{tab:allPLRsI}).
Note that no correction of interstellar extinction has been applied 
(they are of course unnecessary in $W(VI)$).

A break in the slope of the PLR at $\log P\sim 0.4$,
was first found by \citet{Bauer-1999}
and confirmed by later work (\citealt{Soszynski-2010a};
and references therein). 
A break at around $\log P=1$ has been also claimed
especially at optical wavelengths 
(see e.g.~Sandage, Tammann \& Reindl, 2009).
The latter break is less certain than the former and
is expected to be less significant, and possibly absent, in the near-IR
(see table~2 in \citealt{Bono-2010}).
It is also not present in $W(VI)$ \citep{Bono-1999}.

The PLRs in the shortest period range, $0.0<\log P<0.4$,
are steeper than those at longer periods in the SMC, but not in the LMC.
The slope change seems to be a characteristic of Cepheids
in a low-metal environment like the SMC
and dwarf galaxies \citep{Dolphin-2003}.
This break has been also confirmed in the {\it Spitzer} photometry
at the 3.6~$\mu$m and 4.5 $\mu$m bands \citep{Ngeow-2010}.
On the other hand, the slopes for the PLRs at $0.4 < \log P < 1.0$
and $\log P > 1.0$ are nearly the same and the break at $\log P \sim 1$ 
is not evident in these data. This is also consistent with previous studies
involving the infrared PLRs (\citealt{Ngeow-2009}; \citealt{Ngeow-2010};
\citealt{Marengo-2010}).

\begin{figure*}
\begin{minipage}{170mm}
\begin{center}
\includegraphics[clip,width=0.70\hsize]{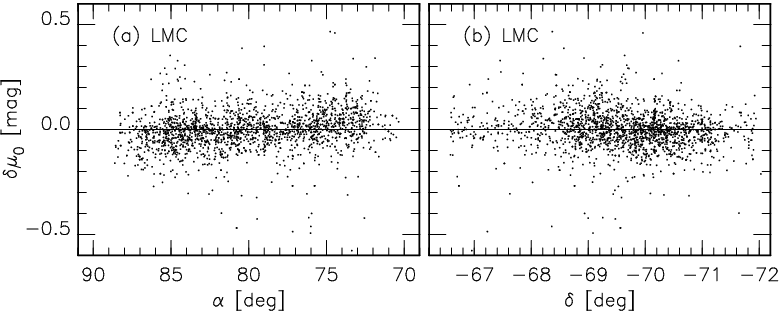}
\includegraphics[clip,width=0.70\hsize]{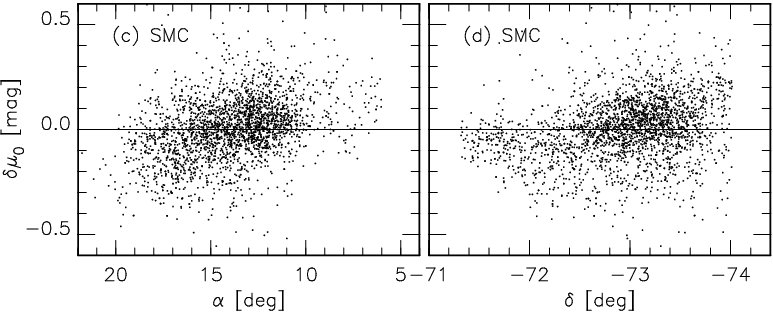}
\end{center}
\caption{The scatter in the distance moduli of individual classical
Cepheids in the LMC, panels (a) and (b), and SMC,
(c) and (d). Refer to the caption of Fig.~\ref{fig:DMscatter2}.
\label{fig:DMscatter}}
\end{minipage}
\end{figure*}

The residuals of individual classical Cepheids from the PLRs
in $W_1(VI)$ for each group are plotted in Fig.~\ref{fig:DMscatter}.
As in Fig.~\ref{fig:DMscatter2}, the distribution for the SMC objects
show a larger scatter than that in the LMC and a clear spatial structure.
One can also notice a small tilt of the distribution in the LMC,
which is barely seen for the T2Cs (Fig.~\ref{fig:DMscatter2}).
The eastern part of the LMC disk is slightly closer to us
than the western part in agreement with previous work
(\citealt{Caldwell-1986}; \citealt{Groenewegen-2000}).

\begin{table*}
\begin{minipage}{0.95\hsize}
\caption{
Period-luminosity relations for classical Cepheids.
Refer to the caption of Table~\ref{tab:PLRs_T2C}.
\label{tab:allPLRsI}}
\begin{center}
\begin{tabular}{ccccccc}
\hline
Band & Group  & Period range & Slope & Zero (@~$\log P_0$) & $\sigma$ & $N$ \\
\hline
$J$&LMC&$0.0<\log P<0.4$&$-3.139\pm 0.124$&$15.522\pm 0.012$~(@~0.3)&0.16&164\\
$J$&LMC&$0.4<\log P<1.0$&$-3.138\pm 0.028$&$14.254\pm 0.005$~(@~0.7)&0.16&1585\\
$J$&LMC&$1.0<\log P<1.7$&$-3.053\pm 0.165$&$12.719\pm 0.020$~(@~1.2)&0.21&119\\
$H$&LMC&$0.0<\log P<0.4$&$-3.235\pm 0.099$&$15.231\pm 0.010$~(@~0.3)&0.13&164\\
$H$&LMC&$0.4<\log P<1.0$&$-3.248\pm 0.022$&$13.926\pm 0.004$~(@~0.7)&0.12&1579\\
$H$&LMC&$1.0<\log P<1.7$&$-3.143\pm 0.139$&$12.336\pm 0.017$~(@~1.2)&0.18&120\\
$\Ks$&LMC&$0.0<\log P<0.4$&$-3.289\pm 0.099$&$15.160\pm 0.010$~(@~0.3)&0.13&164\\
$\Ks$&LMC&$0.4<\log P<1.0$&$-3.284\pm 0.020$&$13.838\pm 0.004$~(@~0.7)&0.11&1584\\
$\Ks$&LMC&$1.0<\log P<1.7$&$-3.197\pm 0.135$&$12.231\pm 0.017$~(@~1.2)&0.17&118\\
$W_1(VI)$&LMC&$0.0<\log P<0.4$&$-3.325\pm 0.078$&$14.906\pm 0.008$~(@~0.3)&0.10& 160\\
$W_1(VI)$&LMC&$0.4<\log P<1.0$&$-3.316\pm 0.015$&$13.573\pm 0.003$~(@~0.7)&0.08&1578\\
$W_1(VI)$&LMC&$1.0<\log P<1.7$&$-3.189\pm 0.163$&$11.916\pm 0.020$~(@~1.2)&0.20& 121\\
$W_2(VI)$&LMC&$0.0<\log P<0.4$&$-3.300\pm 0.080$&$14.978\pm 0.008$~(@~0.3)&0.10& 160\\
$W_2(VI)$&LMC&$0.4<\log P<1.0$&$-3.294\pm 0.015$&$13.654\pm 0.003$~(@~0.7)&0.09&1579\\
$W_2(VI)$&LMC&$1.0<\log P<1.7$&$-3.159\pm 0.160$&$12.010\pm 0.019$~(@~1.2)&0.20& 121\\
\hline
$J$&SMC&$0.0<\log P<0.4$&$-3.381\pm 0.061$&$16.035\pm 0.008$~(@~0.3)&0.23&1590\\
$J$&SMC&$0.4<\log P<1.0$&$-2.992\pm 0.049$&$14.720\pm 0.009$~(@~0.7)&0.22&862\\
$J$&SMC&$1.0<\log P<1.7$&$-3.020\pm 0.128$&$13.171\pm 0.021$~(@~1.2)&0.22&107\\
$H$&SMC&$0.0<\log P<0.4$&$-3.519\pm 0.054$&$15.744\pm 0.007$~(@~0.3)&0.20&1584\\
$H$&SMC&$0.4<\log P<1.0$&$-3.134\pm 0.042$&$14.392\pm 0.007$~(@~0.7)&0.19&854\\
$H$&SMC&$1.0<\log P<1.7$&$-3.110\pm 0.114$&$12.792\pm 0.019$~(@~1.2)&0.19&107\\
$\Ks$&SMC&$0.0<\log P<0.4$&$-3.546\pm 0.055$&$15.683\pm 0.007$~(@~0.3)&0.20&1582\\
$\Ks$&SMC&$0.4<\log P<1.0$&$-3.176\pm 0.041$&$14.315\pm 0.007$~(@~0.7)&0.18&854\\
$\Ks$&SMC&$1.0<\log P<1.7$&$-3.197\pm 0.111$&$12.702\pm 0.018$~(@~1.2)&0.19&108\\
$W_1(VI)$&SMC&$0.0<\log P<0.4$&$-3.543\pm 0.043$&$15.449\pm 0.005$~(@~0.3)&0.16&1596\\
$W_1(VI)$&SMC&$0.4<\log P<1.0$&$-3.318\pm 0.035$&$14.055\pm 0.006$~(@~0.7)&0.15& 859\\
$W_1(VI)$&SMC&$1.0<\log P<1.7$&$-3.283\pm 0.093$&$12.402\pm 0.015$~(@~1.2)&0.16& 108\\
$W_2(VI)$&SMC&$0.0<\log P<0.4$&$-3.521\pm 0.043$&$15.515\pm 0.006$~(@~0.3)&0.16&1596\\
$W_2(VI)$&SMC&$0.4<\log P<1.0$&$-3.286\pm 0.035$&$14.130\pm 0.006$~(@~0.7)&0.16& 859\\
$W_2(VI)$&SMC&$1.0<\log P<1.7$&$-3.250\pm 0.093$&$12.489\pm 0.015$~(@~1.2)&0.16& 108\\
\hline
\end{tabular}
\end{center}
\end{minipage}
\end{table*}

\label{lastpage}
\end{document}